\documentclass[sn-aps,iicol]{sn-jnl}
\usepackage[utf8]{inputenc}
\usepackage[english]{babel}
\usepackage[T1]{fontenc}
\usepackage{lmodern}
\usepackage{microtype}
\usepackage{booktabs}
\usepackage{bm}
\usepackage{graphicx}
\usepackage{amsfonts}
\usepackage{amsmath}
\usepackage{amssymb}
\usepackage{listings}
\usepackage{physics}
\usepackage{subfigure}
\usepackage{fancyhdr}
\usepackage{hyperref}
\usepackage{mathtools}
\usepackage{xcolor}
\usepackage{natbib}

\newcommand{\Gunit}{\mathrm{m}^3\ \mathrm{kg}^{-1}\ \mathrm{s}^{-2}}

\begin{document}

\title[Bayesian analysis of systematic errors in the determination of the constant of gravitation]{Bayesian analysis of systematic errors in the determination of the constant of gravitation}

\author*[1,2]{\fnm{Stefano} \sur{Rinaldi}}\email{stefano.rinaldi@phd.unipi.it}

\author[3,4]{\fnm{Hannah} \sur{Middleton}}

\author[1,2]{\fnm{Walter} \sur{Del~Pozzo}}

\author[5]{\fnm{Jonathan} \sur{Gair}}

\affil[1]{\orgname{Dipartimento di Fisica ``E. Fermi''}, \orgaddress{\street{Università di Pisa}, \city{Pisa}, \postcode{I-56127}, \country{Italy}}}
\affil[2]{\orgname{INFN}, \orgaddress{\street{Sezione di Pisa}, \city{Pisa}, \postcode{I-56127}, \country{Italy}}}
\affil[3]{\orgname{Institute for Gravitational Wave Astronomy \& School of Physics and Astronomy}, \orgaddress{\street{University of Birmingham}, \city{Birmingham}, \postcode{B15 2TT}, \country{United Kingdom}}}
\affil[4]{\orgname{OzGrav-Melbourne \& School of Physics}, \orgaddress{\street{University of Melbourne}, \city{Parkville}, \postcode{Victoria 3010}, \country{Australia}}}
\affil[5]{\orgname{Max Planck Institute for Gravitational Physics (Albert Einstein Institute)}, \orgaddress{\street{Am Mühlenberg 1}, \city{Potsdam}, \postcode{14476}, \country{Germany}}}

\date{\today}

\abstract{
Measurements of the gravitational constant $G$ are notoriously difficult. Individual state-of-the-art experiments have managed to determine the value of $G$ with high precision: although, when considered collectively, the range in the measured values of $G$ far exceeds individual uncertainties, suggesting the presence of unaccounted for systematic effects.

Here, we propose a Bayesian framework to account for the presence of systematic errors in the various measurement of $G$ while proposing a consensus value, following two paths: a parametric approach, based on the Maximum Entropy Principle, and a non-parametric one, the latter being a very flexible approach not committed to any specific functional form. 

With both our methods, we find that the uncertainty on this fundamental constant, once systematics are included, is significantly larger than what quoted in CODATA 2018. Moreover, the morphology of the non-parametric distribution hints towards the presence of several sources of unaccounted for systematics.
In light of this, we recommend a consensus value for the gravitational constant $G = 6.6740^{+0.0015}_{-0.0015} \times 10^{-11}\ \Gunit$.}

\maketitle

\section{Introduction}
The Newtonian constant of gravitation, $G$, is one of the fundamental constants of modern physics. It was the first fundamental constant to be identified and yet it remains one of the least well known, with large disagreement between experimental measurements. Over several decades, huge experimental efforts have tried to determine the value of $G$. Individually, these experiments report relative uncertainties that can be as low as $1.2 \times 10^{-5}$: however different experiments find values of $G$ that can be several standard deviations away from each other. With such a range in measurement, combining results into a single best estimate of $G$ is understandably challenging \citep{speake:2014,wood:2014}.

The current accepted value of $G$ comes from the Committee on Data for Science and Technology (CODATA). CODATA periodically provides a set of self-consistent values of the fundamental constants for use by the scientific and technological communities. The recommended value of $G$ from the CODATA 2010 results is $6.67384(80)\times10^{-11}\ \Gunit$ \citep{mohr:2012}.
After the addition of three more experimental results, the CODATA 2014 recommended value is $6.67408(31)\times 10^{-11}\ \Gunit$ \citep{mohr:2014}. In 2017, a CODATA Special Adjustment \citep{mohr:2018} was released with the purpose of obtaining the best numerical values of the Planck constant $h$, the electron mass $e$, Boltzmann's constant $k$, and Avogadro's number $N_A$: however, the value of $G$ was not updated.

The most recent CODATA 2018 recommendation for $G$ comes from \cite{tiesinga:2021}, where two new experimental results are included and a correction is made to a previously reported value. The current recommended value for $G$ is
\begin{equation*}
G = 6.67430(15)\times 10^{-11}\ \Gunit\,. 
\end{equation*}

Any experiment is affected by noise; the effect of the noise is to induce uncertainty on the quantity of interest. The uncertainty can be \emph{statistical} -- the statistical error is the difference between a value measured in a single experiment and the value averaged over many experiments -- or \emph{systematic}, which is the difference between the averaged value and the true value of the parameter(s) of interest. The main difference among the two classes of uncertainties is that while the statistical error causes a random shift with zero mean of the measured quantity (thus, in principle, the statistical error can be averaged out simply by repeating the experiment a very large number of times), the systematic error for a specific experiment will shift the expected value away from the \emph{true} value of the measured quantity.

The considerable disagreements among different experiments aimed at determining the gravitational constant suggests the presence of an overarching unidentified source of uncontrolled systematic effects leading to such disparate results but, to the best of our knowledge, no other work addresses the presence of systematic errors in a statistical way.

In this paper, we model systematics within the context of Bayesian probability theory. In particular, we will introduce a so-called \emph{hierarchical} model to infer a probability distribution for the unknown systematic errors. In doing so, we will explore several different assumptions, each reflecting a particular choice regarding the nature and magnitude of the errors. The measurements included in this work are the ones listed in \cite{tiesinga:2021}.

The rest of the paper is organised as follows: in Sec.~\ref{measurements} we briefly review the measurements of $G$ used in this work and two existing statistical methods to propose a consensus value. In Sec.~\ref{hier_analysis} we describe the Bayesian hierarchical framework used to estimate $G$. In Sec.~\ref{results} we present our results and finally, in Sec.~\ref{conclusion}, we discuss our findings and conclude with a recommendation on the value of $G$.

\section{Measurements of the gravitational constant}\label{measurements}
The analysis presented in this paper makes use of 16 experimental results dating from 1982 to 2018. Here we briefly review the methods used in each of the 16 experiments. For a comprehensive and detailed review of the measurements of $G$, we refer the interested reader to \cite{speake:2014} or \cite{wood:2014}.

Following the approach taken by Cavendish in 1797–1798, the majority of experiments listed in \cite{tiesinga:2021} involve precision measurements of a torsion balance. Free deflection was used in BIPM-01 and BIPM-14 \citep{quinn:2001,quinn:2013} as well as electrostatic compensation (see below). Time-of-swing experiments (NIST-82, TR\&D-97, LANL-97, HUST-05, HUST-09, UCI-14 and HUST$_{\mathrm{T}}$-18) instead measure the change in oscillation period of the torsion balance with different source mass positioning \citep{luther:1982,karagioz:1996,bagley:1997,hu:2005,luo:2009,tu:2010,newman:2014,li:2018}.
A third variation on the torsion balance uses electrostatic compensation (BIPM-01, MSL-03 and BIPM-14) \citep{quinn:2001,armstrong:2003,quinn:2013}. The gravitational torque on the test masses is balanced by an electrostatic torque so that they do not rotate. For UWash-00 and HUST$_{\mathrm{A}}$-18, the torsion balance is rotated on a turntable and feedback is used to change the rotation rate so that the fibre twist is minimised and the angular acceleration of the turntable is equal to the gravitational angular acceleration of the balance \citep{gundlach:2000,li:2018}.

Four experiments listed in \cite{tiesinga:2021} do not use a torsion balance method. UWup-02 uses a microwave Fabry-Perot interferometer whose resonance frequency is influenced by the placement of source masses behind each of the reflectors \citep{kleinevoss:2002}. Similarly, JILA-18 uses a laser Fabry-Perot interferometer to measure the spacing between the test masses of a double pendulum as the positions of source masses around it are changed \citep{parks:2010,parks:2019}. UZur-06 uses a beam balance to weigh test masses in the presence of movable source masses \citep{schlamminger:2006}. LENS-14 uses atom interferometry to measure how a source mass influences the atom's acceleration \citep{prevedelli:2014,rosi:2014}.

\cite{tiesinga:2021} reports the value of $G$ and one-sigma uncertainties $\sigma$ for each experiment. The range of the measured values is $\approx 0.0037 \times 10^{-11}\ \Gunit$: however, the largest individual one-sigma uncertainty is $0.00099\times 10^{-11}\ \Gunit$, from LENS-14. On the other hand, other measurements report uncertainties as small as $\approx 0.00008\times 10^{-11}\ \Gunit$ (HUST-18). 

\begin{figure}
    \centering
    \includegraphics[width = 0.8\columnwidth]{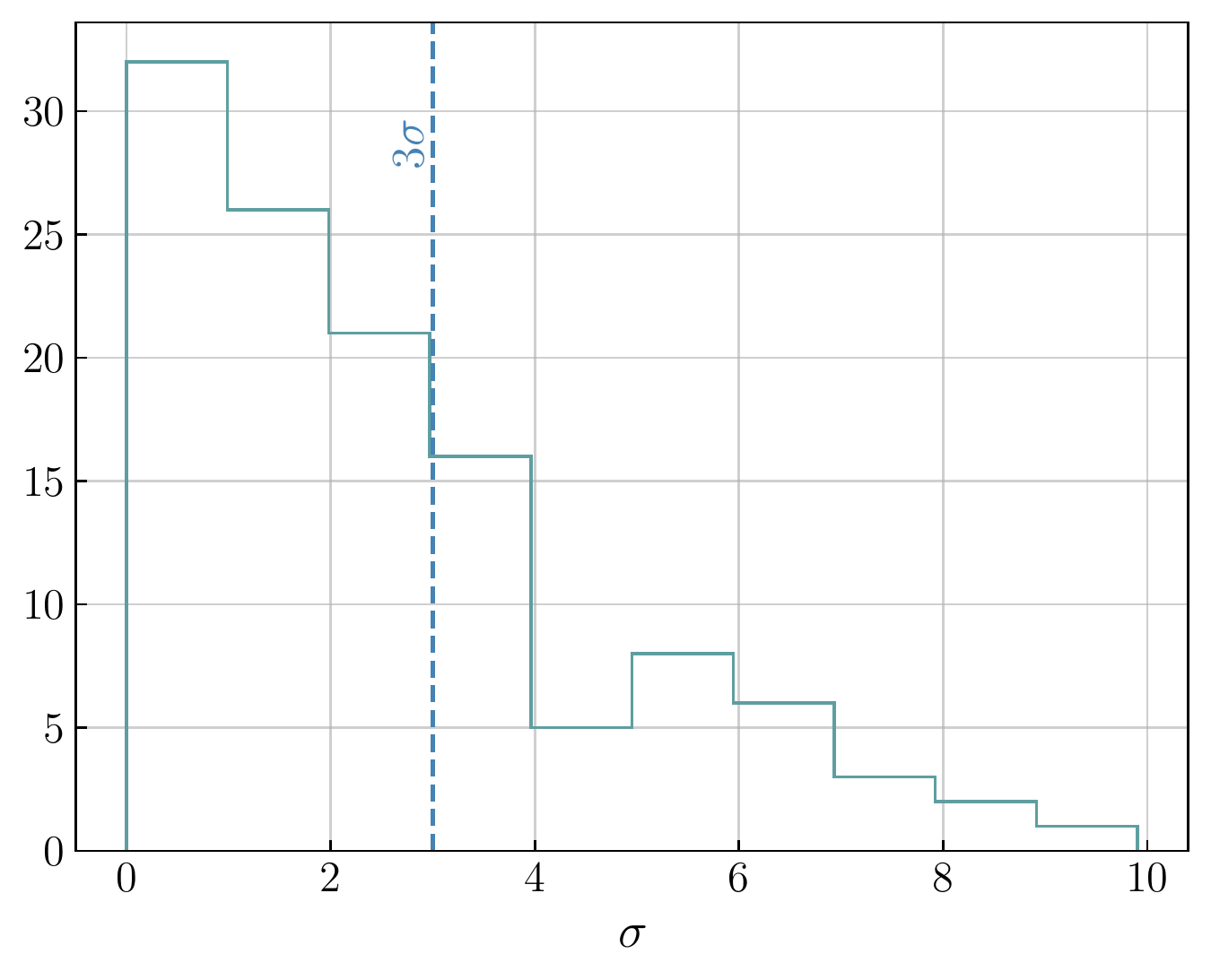}
    \caption{Histogram of $\sigma$-levels for all the possible pairs of experiments included in \cite{tiesinga:2021}. Assuming a threshold of $3\sigma$ for a pair of measurements to be in agreement with each other, 41 pairs out of 120 fall beyond this threshold.}
    \label{fig:sigmas}
\end{figure}

Most importantly, individual observations are often inconsistent with others within their stated measurement uncertainties. If we assume that measurements are inconsistent above the $3\sigma$ level, we find that $34\%$ of all the possible pairs of experiments are inconsistent among themselves (see Figure~\ref{fig:sigmas}).

Recommending a single value from this variety of measurements is understandably difficult.

\subsection{Existing statistical frameworks}
This is not the first work that tries to reconcile the plethora of different values for G into a single, recommended value. Most of the previous effort, however, is devoted to the identification of potential sources of discrepancy among different experiments, ranging from systematic errors in the measurement apparatuses to the potential presence of an unknown oscillatory factor affecting the measurement process over time or accounting for inaccuracies in Newtonian theory \citep{pitkin:2015,anderson:2015}.

Among the few works that tries to address the discrepancy among different experiments from a statistical point of view, we outline here two existing frameworks to propose a consensus value for $G$. The first is the one used by \cite{tiesinga:2021}, whereas the second is introduced in \cite{merkatas:2019}. Neither of them, however, proposes a framework to account for systematic errors. Here, we briefly review these techniques and in Section~\ref{hier_analysis} we present a new statistical method to account for the presence of systematics in the consensus value.

\subsubsection{Tiesinga \emph{et al.}: least-squares procedure}
The value proposed in \cite{tiesinga:2021}, as well as the ones from previous CODATA recommendations, is obtained using a least-squares procedure. In particular, according to \cite{wood:2014}, having $n$ different measurements $\mathbf{y} = \{y_1,\ldots, y_n\}$ of an unknown quantity $\bar{y}$ with covariance matrix $\mathbf{C}$, minimising the quantity
\begin{equation}
    \chi^2 = (\mathbf{y}-\bar{y})\mathbf{C}^{-1} (\mathbf{y}-\bar{y})^{\mathrm{T}}
\end{equation}
with respect to $\bar{y}$ leads to the variance-weighted mean of the measurement and its uncertainty.

This method, however, relies on a fundamental assumption: the provided uncertainties have to be statistical in nature rather than systematic. This procedure weights the different measurements according to their precision, trusting more the measurements with the smallest associated error, which is reasonable under the assumption that all the measurements are in agreement with the same \emph{true} value.

In presence of systematic errors, however, there is no reason to believe that the most \emph{precise} measurement is also the most \emph{accurate}. As stated above, the presence of systematic errors shifts the expected value of the affected measurement: thus, the line of reasoning in which we favour the measurement obtained with a very precise experiment might end up being biased.

\subsubsection{Merkatas \emph{et al.}: shades of dark uncertainty}
This work \cite{merkatas:2019} suggests that the uncertainty associated with each of the different measurements of the gravitational constant does not account for \emph{all} the statistical uncertainty and that there could be several different latent sources of statistical uncertainty. 

The idea is that different experiments are affected by these \emph{shades of dark uncertainty} and that the same source can be shared among different experiments, providing random amounts of additional statistical uncertainty. They propose a Bayesian framework to infer the magnitude of these additional uncertainties required to reach consensus among measurements and then propose a value for $G$. Their work, however, does not consider the possibility of systematic errors being present.

This approach, in our view, makes sense while grouping experiments that are correlated in some sense (e.g. using the same methodology or being performed by the same people): this way, there is room to believe that uncertainties in similar experiments could have been similarly underestimated.

\section{Bayesian hierarchical analysis}\label{hier_analysis} 
In this section we propose a method to combine different measurements of $G$ by employing a hierarchical framework based on Bayesian inference as a way of marginalising over the unknown systematic effects.

Let us begin by defining the value of the gravitational constant as $G$; we wish to determine $G$ given the ensemble of $N$ experiments $\mathbf{D}=\{D_1,\ldots,D_N\}$ and a model $H$.

The whole idea of \emph{setting up an experiment}, given the presence of a different systematic error in every measurement, can be represented as follows: every experiment, in presence of systematics, will measure an \emph{experiment value} $G_i$, which is not the true value of the constant of gravitation $G$. This value is a realisation of a stochastic process governing the source of systematic errors, being drawn from $p(G\rvert\theta)$:

\begin{equation}
    G_i \sim p(G \rvert \theta)\,.
\end{equation}
$\theta$, here, represents the parameters of our model for the distribution of systematic errors.

Each individual experiment $i$, in turn, will result in a probability distribution for its own $G_i$. Hierarchically combining different measurements, therefore, allows us to characterise the probability distribution of systematic errors. This distribution, at the same time, acts as a probability distribution for $G$, characterising the probability of deviating from the unknown true value: therefore, at the end of this work, we will recommend a value for $G$ based on this probability density.

The application of this population study-like approach is possible thanks to the independence of all the systematic errors at play\footnote{For most of the measurements included in this work, this is a safe assumption. Three pairs of these experiments, however, are correlated (see caption of Table XXIX in \cite{tiesinga:2021}). Given the fact that these correlations are low, with the highest correlation coefficient being $r$(NIST-82,NANL-97) = 0.351, we opted to neglect these correlations.} under the assumption that, although two different experiments may share the same source of systematics (e.g. using the same experimental setup), the magnitude of the systematic is different for each of them.

The probability distribution $p(G\rvert\theta)$ is determined by its functional form and by a set of parameters $\theta$. If one wants to reconstruct the systematic error distribution -- which means assuming a functional form for $p(G\rvert\theta)$ and inferring its parameters $\theta$ -- and therefore give a probability distribution for $G$, the data to use in such inference are the experiment values $\mathbf{G} = \{G_1,\ldots G_N\}$. Unfortunately, due to the presence of statistical uncertainty, these values are unknown, since the $G_i$s are the values that would be measured by experiments in absence of statistical error: every experiment, implicitly, gives a probability distribution for this quantity while reporting a value with an associated error.
Having at hand only the $N$ posterior distributions $\mathbf{D}$ provided by our experiments, we need to combine the experiment outcomes $\mathbf{D}$ in a hierarchical fashion.

Within the Bayesian framework and under the assumption of a functional form for the systematic error distribution\footnote{This assumption is included in the hypothesis $H$.}, the inference is completely described by the posterior distribution for $\theta$:
\begin{equation}
    p(\theta\rvert\mathbf{D},H) = \frac{p(\mathbf{D}\rvert\theta,H)p(\theta\rvert H)}{p(\mathbf{D}\rvert H)}\,,
\end{equation}
where $p(\theta\rvert H)$ is the prior probability distribution for $\theta$, describing our \emph{a priori} expectation for its value.
The likelihood function $p(\mathbf{D}\rvert\theta,H)$ is known only conditioned on the knowledge of $\mathbf{G}$. Marginalising over this quantity, we get
\begin{equation}
    p(\theta\rvert\mathbf{D},H) = \frac{p(\theta\rvert H)\int p(\mathbf{D}\rvert\theta, \mathbf{G},H) p(\mathbf{G}\rvert\theta,H) \dd \mathbf{G}}{p(\mathbf{D}\rvert H)}\,.
\end{equation}
Here $p(\mathbf{G}\rvert\theta,H)$ represents the systematic error distribution. Under the assumption of statistical independence of each experiment, can be factorised into the product of probabilities:
\begin{equation}
    p(\mathbf{G}\rvert\theta,H) = \prod_i^N p(G_i\rvert\theta,H)\,.
\end{equation}
The denominator $p(\mathbf{D}\rvert H)$ is the so-called \emph{evidence}, which is given by the integral over all the parameters characterising the statistical model induced by the hypothesis $H$.

\subsection{Likelihood function}
The likelihood function $p(\mathbf{D}\rvert\theta,\mathbf{G},H)$ describes, in fact, the likelihood of observing the available data given a specific value for the parameters that we want to infer. Making use, once again, of the assumption of statistical independence of each experiment and of the fact that each $D_i$ is independent of $G_j$ for $j\neq i$, the likelihood factorises into the product of individual likelihoods:
\begin{equation}
    p(\mathbf{D}\rvert\theta,\mathbf{G}, H) = \prod_i^N p(D_i\rvert G_i,H)\,.
\end{equation}
Once the \emph{experiment value} of $G$, $G_i$, is known, the posterior distribution for each experiment does not depend on the values of the parameters $\theta$, since these describes only the systematic error distribution: this is a consequence of the fact that systematic errors cannot be removed or accounted for \emph{a posteriori}.

Every experiment implicitly gives a posterior distribution for $G_i$, hence $p(G_i\rvert D_i,H)$: making use of the Bayes' theorem, we get
\begin{equation}
    p(\mathbf{D}\rvert\theta,\mathbf{G}, H) = \prod_i^N \frac{p(G_i\rvert D_i,H)p(D_i\rvert H)}{p(G_i\rvert H)}\,.
\end{equation}
$p(G_i\rvert H)$ is the prior on each $G_i$, which we take uniform between $G_{\mathrm{min}} = 6.668\times 10^{-11}\ \Gunit$ and $G_{\mathrm{max}} = 6.678\times 10^{-11}\ \Gunit$, and $p(D_i\rvert H)$ is the evidence for the single experiment outcome.

Our framework needs to include a functional form for these $N$ posterior distributions.
Given that the only information we have available are the central value $\hat{G}_i$ and the uncertainty $\sigma_i$ around it -- therefore $D_i = \{\hat{G_i}, \sigma_i\}$ -- following the Maximum Entropy Principle (MEP) \citep{jaynes:2003} we assume a Gaussian distribution\footnote{The MEP states that the probability distribution that maximise the information entropy making use of the least amount of information or, in some late sense, the most conservative choice knowing only the expected value and the variance is the Gaussian distribution.}. Under this assumption, the likelihood for each measurement reads
\begin{equation}\label{likelihood_ff}
    p(G_i\rvert D_i,H) \propto \exp[-\frac{1}{2} \qty(\frac{G_i - \hat{G}_i}{\sigma_i})^2]\,.
\end{equation}

\subsection{Systematic effects modelling}
In order to reconstruct the probability distribution $p(G\rvert\theta)$, we need to assume a model for this distribution. Here, we propose two different models, based on different assumptions.
\subsubsection{Maximum Entropy Principle: Gaussian distribution}
We model the effect of the unknown systematic errors as follows: since we consider only the dispersion of systematics, we once again appeal to the MEP to choose the probability distribution $p(G\rvert\theta)$. 
Given that we want to give an expected value and an uncertainty for $G$, the distribution is taken to be a Gaussian distribution with mean $\hat{G}$ and unknown standard deviation $\Sigma$, therefore $\theta = \{\hat{G}, \Sigma\}$.

Under this assumption, we can write:
\begin{equation}
    p(G_i\rvert\theta, H) = \exp[-\frac{1}{2}\qty(\frac{G_i-\hat{G}}{\Sigma})^2]\,.
\end{equation}

The assumption of a Gaussian distribution both for $p(G_i\rvert D_i)$ and $p(G_i\rvert\theta)$ is particularly useful, since it is possible to marginalise over $G_i$ analytically. In fact, making use of the fact that the integrals are independent,
\begin{multline}
    \prod_i^N\int \mathcal{N}(G_i\rvert\hat{G},\sigma_i)\mathcal{N}(G_i\rvert\hat{G},\Sigma) \dd G_i \\=\prod_i^N \mathcal{N}\qty(\hat{G}_i\bigg\rvert\hat{G},\sqrt{\sigma_i^2+\Sigma^2})\,,
\end{multline}
where we denoted with $\mathcal{N}(\cdot\rvert\mu,\sigma)$ the Gaussian distribution.

The prior $p(\theta\rvert H)$ is composed by the prior on $\hat{G}$, which we take uniform between $G_{\mathrm{min}}$ and $G_{\mathrm{max}}$, and the prior on $\Sigma$. We will consider several possible choices for this distribution, following some of the prescriptions discussed in \citep{gelman:2006}:
\begin{itemize}
    \item UN: a uniform distribution for $\Sigma$. Using this prior probability distribution means assuming that we have no information at all regarding the value of the systematic error;
    \item JF: a uniform distribution over $\log\Sigma$.  This is the so- called Jeffreys' prior, corresponding to the assumption that we have no information about the value of the order of magnitude of $\Sigma$. A change of variable shows that the probability density function for $\Sigma$ is proportional to $1/\Sigma$, hence reflecting the expectation that the systematic errors are small;
    \item IG: an Inverse Gamma distribution
    \begin{equation}
        p(\Sigma\rvert\alpha,\beta) = \frac{\beta^\alpha}{\Gamma(\alpha)}\Sigma^{-(\alpha+1)}\exp[-\frac{\beta}{\Sigma^2}]\,,
    \end{equation}
    where $\alpha > 0$ and $\beta > 0$ are called the \emph{shape} and \emph{scale} parameters that determine the morphology of the Inverse Gamma distribution, and $\Gamma(\alpha)$ is the complete Gamma function.
    We infer $\alpha$ and $\beta$ from the experimental values, assigning uniform priors between 0 and 100. The Inverse Gamma distribution is \emph{conjugate} to the Gaussian distribution. This guarantees that the posterior on $\Sigma$ is still an Inverse gamma distribution.
\end{itemize}

\subsubsection{Non-parametric reconstruction: (H)DPGMM}
The second model we use is (H)DPGMM, a non-parametric model introduced in \cite{rinaldi:2022:hdpgmm}. In what follows, we will give a brief overview of the model, referring the interested reader to the relevant papers for more details.
Bayesian non-parametric methods are powerful tools that allow us to perform an inference without committing to any specific model prescription. This results in an extreme flexibility when it comes to modelling unknown distributions: all the information that is encoded in the inferred distribution is extracted from the data themselves.

In particular, this model relies on the Dirichlet Process Gaussian Mixture Model \citep{escobar:1995} or DPGMM, an infinite weighted sum of Gaussian distributions with a Dirichlet Process \citep{ferguson:1973} as prior distribution on weights, to approximate the unknown probability distribution:
\begin{equation}
    p(x) \approx \sum^{\infty}_i w_i \mathcal{N}(x\rvert\mu_i,\sigma_i)\,.
\end{equation}
The \emph{standard} DPGMM is used to reconstruct an \emph{outer} probability distribution when samples $\mathbf{x} = \{x_1,\ldots,x_N\}$ from the unknown distribution $p(x)$ are available.
This is not always the case: there are situations, like the mass function inference described in \cite{rinaldi:2022:hdpgmm}, in which we do not have direct access to samples, but rather we have $N$ sets of \emph{inner} samples drawn from the $N$ posterior distributions (\emph{inner distributions}) for each sample $x_i$. 

To infer the outer distribution having at hand only the $N$ sets of inner posterior samples, one needs to specify a model for both the outer distribution and for the $N$ inner posterior distributions: (H)DPGMM models both the inner and the outer distributions as DPGMM, linking them in a hierarchical fashion.

This is a very similar situation to the one we are addressing in this paper. $p(G)$ is the outer distribution and $p(G_i\rvert D_i)$ are the $N$ posterior distributions we want to use to infer $p(G)$. In general, in order to apply (H)DPGMM, we would need to approximate $p(G_i\rvert D_i)$ with a weighted sum of Gaussian distributions: however, we can interpret the likelihood~\eqref{likelihood_ff} as a DPGMM with a single component with $w_i = 1$, whereas every other Gaussian component has $w_j = 0$, and use it as a very simple non-parametric reconstruction.

In this case, the parameter vector $\theta = \{\mathbf{w}, \boldsymbol\mu,\boldsymbol\sigma\}$ is composed of a vector of relative weights, a vector of means and a vector of standard deviations.
The length of these vectors is, \emph{a priori}, not limited.

The outcome of such a model, applied to the problem we are dealing with, is a phenomenological distribution for the gravitational constant $G$.

\subsection{Inference}
We proceed now to specify how the parameters $\theta$ for each hypothesis are inferred.

The expression for the posterior distribution, under the MEP hypothesis, becomes
\begin{equation}\label{post_gaussian}
    p(\theta\rvert\mathbf{D}, H) \propto p(\theta\rvert H) \prod_i^N \mathcal{N}\qty(\hat{G}_i\bigg\rvert\hat{G},\sqrt{\sigma_i^2+\Sigma^2})\,.
\end{equation}
For each of the three different prior prescriptions for $\Sigma$ (UN, JF and IG), we generate samples from Eq.~\eqref{post_gaussian} using a nested sampling algorithm \citep{skilling:2006}, \textsc{CPNest} \citep{cpnest}.

For the non-parametric hypothesis, (H)DPGMM, we explore the posterior distribution drawing different realisations for $p(G)$ using \textsc{figaro}, a Gibbs sampler presented in \cite{rinaldi:2022:figaro}.

\section{Results}\label{results}
We summarise here our findings for each of the systematic error models considered in this work.

\begin{figure*}
    \centering
    \includegraphics[width = 1.5\columnwidth]{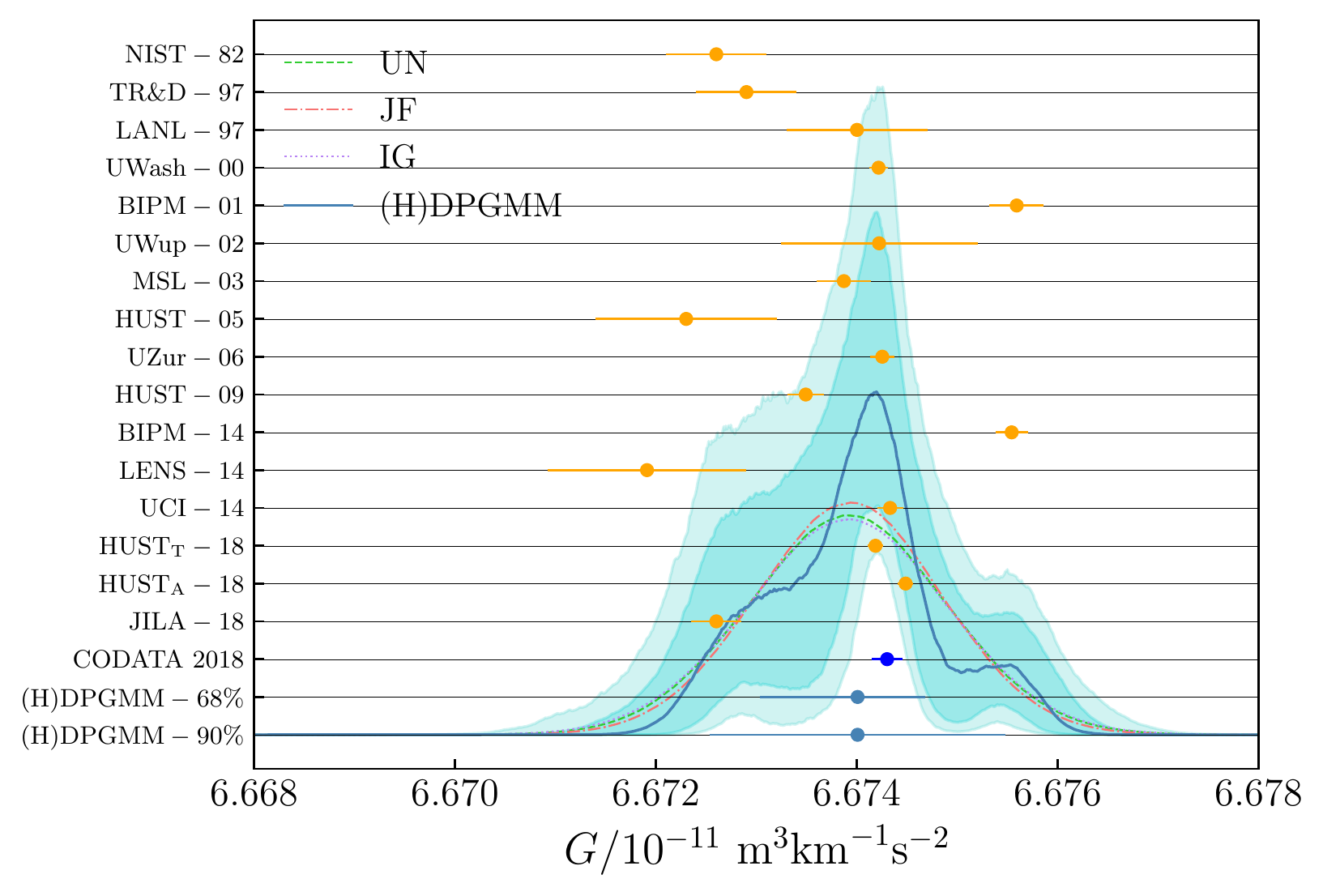}
    \caption{Median posterior distribution for $G$ under the models presented in this paper. For the (H)DPGMM reconstruction, the solid blue line represents the median and the shaded regions indicates the $68\%$ (dark turquoise) and $90\%$ (light turquoise) credible regions for the posterior on $G$. As a comparison, we report the experimental values and their standard error (orange symbols), as well as the \cite{tiesinga:2021} (CODATA 2018) recommended value (blue symbol) and our most probable value for $G$ and $68\%$ and $90\%$ (conservative) credible intervals for (H)DPGMM.}
    \label{fig:posteriors}
\end{figure*}

Posterior distributions for $G$ under the UN, JF, IG and (H)DPGMM hypothesis are shown in Figure~\ref{fig:posteriors}.
The shape of the non-parametric distribution for $G$ is very different from the shape of the simple Gaussian distribution assumed under the MEP hypothesis: in fact, three different modes are clearly distinguishable.  We retain this fact as qualitative evidence of the presence of uncontrolled systematics.

In the following, we summarise the main findings for the proposed models by discussing the inferred posterior distribution, reporting median and $68\%$ credible interval from the median posterior:
\begin{itemize}
    \item UN: $G = 6.6739(10) \times 10^{-11}\ \Gunit\,;$
    \item JF: $G = 6.6739(9) \times 10^{-11}\ \Gunit\,;$
    \item IG: $G = 6.6739(10) \times 10^{-11}\ \Gunit\,.$
\end{itemize}
Concerning (H)DPGMM, the median posterior distribution for $G$ under this hypothesis gives median and $68\%$ credible interval
\begin{equation*}
    G = 6.6740^{+0.0007}_{-0.0009} \times 10^{-11}\ \Gunit\,,
\end{equation*}
and median and $90\%$ credible interval
\begin{equation*}
    G = 6.6740^{+0.0015}_{-0.0015} \times 10^{-11}\ \Gunit\,.
\end{equation*}
The presence of three different modes in the non-parametric reconstruction could hint towards the interpretation of \emph{families} of systematic effects, similar to the founding idea of \cite{merkatas:2019}: before commenting on this, however, we want to make clear that an extensive discussion of the systematics that might affect the individual experiments is well beyond our area of expertise. Therefore, the following discussion must be taken as heuristic and driven by statistical considerations only: before claiming that two or more experiments are affected by the same, or at least similar, systematics, it is necessary a dedicated study on the potential sources of such systematic errors.

These three modes might suggest that at least three (or two, if we assume that one of these modes is free of systematics) different effects are at play. While it is not possible to tell which (if any) of the three modes is unaffected by systematics, we note that the rightmost mode contains two measurements only, BIPM-01 and BIPM-14. These two experiments share both the same methodology and the same group, making plausible (not \emph{likely}) for their results to be affected by the same source of systematics\footnote{For intellectual honesty, we note that it might also be possible to suggest, making use of the same line of reasoning, that these two are the only experiments unaffected by systematics.}.

An alternative -- but incorrect -- interpretation of these results might be to use the mean parameter of the Gaussian distribution $\hat{G}$ from the MEP model as the true value of the gravitational constant $G$. The inferred value is $\hat{G} = 6.6739 \pm 0.0003\times 10^{-11}\ \Gunit$, estimated via Monte Carlo sampling. This value is similar to the method and result presented in \cite{merkatas:2019}, in which the authors address the same issue by proposing an additional, latent source of uncertainty.

Interpreting this quantity, which we report for completeness, as the true value for the gravitational constant, in this framework, is conceptually incorrect: $\hat{G}$ is a parameter of the posterior distribution for $G$. For a Gaussian distribution, the median coincides with the mean parameter: the uncertainty on the inferred mean parameter is, however, in general much smaller than the variance of the distribution, leading to an underestimation of the uncertainty on $G$.

\section{Conclusions}\label{conclusion}
In this paper, we proposed two different models to reconstruct the probability distribution $p(G)$ assuming the presence of systematic effects. We found that, although the numerical values for the gravitational constant are very similar among the two models, the functional form reconstructed by the non-parametric one is morphologically different from the Gaussian distribution that arises from the MEP hypothesis. 

This suggests that the systematic effects at work behind the experiments we considered are not under control: although some of these measurements are extremely \emph{precise}, they are not very \emph{accurate}. Therefore, further studies are required both to understand the systematics that affect these experiments and to pinpoint the value of the gravitational constant. Such studies are already taking place, as described in \cite{rothleiner:2017} and references therein or in \cite{schlamminger:2022}.

In light of our investigations, we find that the latest CODATA recommended value is heavily underestimating the actual uncertainty on $G$. Hence, although this is not the purpose of this paper, we think that the best value to adopt for $G$ is the most conservative we find under the most general assumptions, the one from the (H)DPGMM model:
\begin{equation*}
    G = 6.6740^{+0.0015}_{-0.0015} \times 10^{-11}\ \Gunit\,.
\end{equation*}
\section*{Acknowledgements}
The authors are grateful to Clive~C.~Speake, Ilya~Mandel, and Alberto~Vecchio for the useful discussions. We thank also the authors of \cite{tiesinga:2021} for the comments.

\section*{Statements and declarations}

\subsection*{Funding}
H.M. acknowledges support of the Australian Research Council Centre of Excellence for Gravitational Wave Discovery (OzGrav) (project number CE170100004) and support of the UK Space Agency, Grant No. ST/V002813/1.

\subsection*{Code availability}
All the results and figures presented in this paper can be reproduced using the code available at \url{https://github.com/sterinaldi/fund_const}.\\
\textsc{figaro} is publicly available at \url{https://github.com/sterinaldi/figaro}.

\subsection*{Data availability}
This manuscript has no associated data.

\subsection*{Competing interests}
All authors certify that they have no affiliations with or involvement in any organization or entity with any financial interest or non-financial interest in the subject matter or materials discussed in this manuscript.

\bibliography{bibliography.bib}
\end{document}